# Nanoscale inhomogeneity of charge density waves dynamics in La$_{2-x}$Sr$_x$NiO$_4$

Gaetano Campi[1], Antonio Bianconi[1,2], Boby Joseph[2,3], Shrawan Kr Mishra[4,5], Leonard Müller[6], Alexey Zozulya[6*], Agustinus Agung Nugroho[7], Sujoy Roy[5], Michael Sprung[6], Alessandro Ricci[2,6**]

[1] Institute of Crystallography, CNR, via Salaria Km 29.300, Monterotondo Roma, 00015, Italy
[2] Rome International Center Materials Science Superstripes (RICMASS), Via dei Sabelli 119A, 00185 Roma, Italy
[3] Elettra Sincrotrone Trieste. Strada Statale 14 - km 163.5, AREA Science Park, 34149 Basovizza, Trieste, Italy
[4] School of Materials Science and Technology, Indian Institute of Technology (BHU), Varanasi 221005, India.
[5] Advanced Light Source, Lawrence Berkeley National Laboratory, Berkeley, California 94720, USA
[6] Deutsches Elektronen-Synchrotron DESY, Notkestraße 85, 22607 Hamburg, Germany
[7] Faculty of Mathematics and Natural Sciences, Institut Teknologi Bandung, Jl. Ganesha 10 Bandung, 40132, Indonesia

* present address European XFEL, Holzkoppel 4, 22869 Schenefeld, Germany.
**present address Duferco Corporate Innovation, Via Trevano 2A, 6900 Lugano, Switzerland

**While stripe phases with broken rotational symmetry of charge density appear in many complex correlated systems, the heterogeneity of spatial ordering and dynamics remains elusive. This missing info is at the heart of understanding the structure and function relation in quantum complex materials. We focus here on the spatial heterogeneity of the motion of charge density wave (CDW) at nanoscale in the archetypal case of La$_{2-x}$Sr$_x$NiO$_{4+y}$ perovskite at low temperature. We report compelling evidence that the unconventional increasing motion of CDW at T < 50K is related with the decreasing of its correlation length using resonant soft X-ray photon correlation spectroscopy (XPCS). The key result of this work is the direct visualization of nanoscale inhomogeneity of CDW relaxation dynamics by scanning micro X-ray diffraction (SµXRD) showing a nanoscale landscape of percolating short range *dynamic* CDW puddles competing with *large quasi-static* CDW puddles giving rise to a novel form of nanoscale phase separation of the incommensurate stripes order landscape.**

Inhomogeneity of charge density waves (CDW) in doped perovskites based on copper [1-14] and nickel [15-35] is today of high increasing interest for understanding the intrinsic disorder in quantum complex matter [36-49] which control emerging quantum functionalities (e.g. high temperature superconductivity, metal-insulator transitions). Novel experimental methods show that nanoscale phase separation in charge, orbital, spin, and lattice degrees of freedom at multiple length scales gives rise to rich landscapes in quantum complex materials. Nowadays, scanning techniques using synchrotron X-ray beams focused down to the nanometer scale have allowed us to visualize unique spatial complexity in quantum materials [41-49]. As relevant examples, we mention here the imaging of intrinsic phase separation in high temperature cuprate superconductors forming a network of nanoscale oxygen rich patches, interspersed with oxygen depleted regions in $HgBa_2CuO_{4+y}$ [10], $La_2CuO_{4+y}$ [44-45] and $YBa_2Cu_3O_{6+y}$ [47,48]. Such an intrinsic phase separation has been observed for iron-based chalcogenide superconductors $A_xFe_{2-y}Se_2$ and Sc doped $Mg_{1-x}Sc_xB_2$ [43], too. Moreover, scanning micro X-ray diffraction (SµXRD) has succeeded to visualize the nanoscopic phase separation with the formation of SDW puddles [28].

Here, we focus our studies on $La_{2-x}Sr_xNiO_4$, a 2D doped Mott insulator, which is isostructural with the superconducting cuprate $La_{2-x}Sr_xCuO_4$. $La_{2-x}Sr_xNiO_4$ shows only localized holes with $Ni3d^8\underline{L}$ configuration, where the ligand hole $\underline{L}$ is a $O(2p^5)$ localized charge [23-25]. This system is known to show SDW as well as CDW at low temperature. The dynamics of the incommensurate SDW has been recently investigated by X-ray Photon Correlation Spectroscopy (XPCS) [20]. The spin-stripe order has been found to be spatially and temporally destabilized at low temperature, resembling the anomalous decay of the stripe order in cuprates, which is generally ascribed to the competing onset of superconductivity. Since superconductivity is absent in this nickelate the atypical low temperature behavior has been explained with an energy barrier for thermally activated lattice fluctuations at nanoscale. [21,22].

Here, we have studied the spatial distribution of the CDW fluctuations rate in $La_{2-x}Sr_xNiO_{4+y}$ by combining resonant soft XPCS with SµXRD. SµXRD and XPCS provide a visualization of the inhomogeneous landscape and the rate of fluctuations of CDW textures, respectively. Using XPCS we have found that the CDW fluctuations rate increases at low temperatures, as it occurs for the spin order [20]. This dynamic rate has been found to be correlated with the CDW coherence length. The spatial map of CDW correlation length has been reconstructed by SµXRD with micrometer resolution. The results show a phase separation between i) *dynamic* CDW puddles with short coherence length and faster fluctuation rate and ii) *quasi-static* CDW puddles with long coherence length and slower fluctuation rate. In this way, we have been able to visualize the spatially inhomogeneous dynamics of CDW in $La_{2-x}Sr_xNiO_4$, which is a key step to shed light on the inhomogeneous spatial distribution of charge density wave puddles in complex quantum materials.

In $La_{1-x}Sr_xNiO_4$ [15] for 0.33>x>0.25 the CDW and SDW extend in real space diagonally to the Ni-O bond directions, along the orthorhombic unit cell. For samples with tetragonal symmetry, as the one studied here, the stripes order itself breaks the rotational symmetry of the *ab*-plane and therefore stripes with two different orientations show up related by a 90-degree rotation around *c*-axis. It has been found by neutron diffraction that incommensurate SDW at the lowest momentum transfer occur at wave vectors (1-ε, 0, 0), where ε is a temperature dependent incommensurability value. CDW reflections can be detected separately in the k-space at (1-$q_h$, 0, L), where $q_h$=2ε and L is odd. In this work we have investigated a La$_{1-x}$Sr$_x$NiO$_4$ sample with a doping level of nominal Sr concentration x=0.28, In previous X-ray diffraction studies on CDW in superconducting cuprates, it has been found that both the correlation length as well as the reflection intensity decrease as superconductivity arises for T<T$_c$. These results have been interpreted as evidence for competition between superconductivity and CDW order. However, a recent study [20] on SDW order in the same insulating nickelate as studied here, shows that the SDW order decreases below 70K while there is no onset of superconductivity. In this context, it is also important to know the behavior of the CDW, in particular at low temperatures. Thus, we have measured the low temperature time fluctuations of the CDW peak with wavevector **q$_{CDW}$** = (1-2ε)**a\*+c\***, by using resonant XPCS, illuminating the sample by soft coherent X-rays with the X-ray energy tuned to the Ni 2p → 3d (L$_3$) resonance. The sketched experimental layout is shown in **Figure 1a**. XPCS measurements have been performed at the Advanced Light Source (ALS) at Lawrence Berkeley National Laboratory. We collected time series of coherent X ray diffraction images of the CDW peak at different temperatures (see Figure 2b and methods). Thanks to the coherence of the X-ray beam, the resonant CDW peak shows many speckles due to the disorder of the charge order domains in the coherently illuminated sample volume. The temporal evolution of the speckles reveals the dynamical behavior within the CDW order. To quantify this evolution, we have determined the intensity autocorrelation function g$_2$ which leads to the modulus of the intermediate scattering function |F(q, t)| through

$$g_2\left(t\right) = \frac{\langle I(\tau)I(\tau+t)\rangle_\tau}{\langle I(\tau)\rangle_\tau^2} = 1 + A|F(q,t)|^2 \qquad (1)$$

where <…>$_\tau$ denotes the integration over the whole set of frames recorded for one temperature. We restrict our analysis to the central part of the Bragg peak, indicated by the black square, where peak intensity is high and more stable. We report no significant indications for different temporal behavior in different regions of the peak. We collected several time-series in the temperature range from 35 to 105 K; the square of |F(q,t)| is shown for each of these temperatures in Figure 1c. All curves show a characteristic

exponential decay. We can distinguish two temperature ranges: in the first one (left panel), heating the sample up to 65K, the characteristic decay time increases (i.e. the curves shift to the right), while in the second regime, for T>65K, the dynamic behavior is inverted and the characteristic decay time decreases. In order to quantify this behavior, we fitted the autocorrelation function by a stretched exponential Kohlrausch-Williams-Watts (KWW) model:

$$|F(t)|^2 = e^{-(t/\tau)^\beta} \qquad (2)$$

where $\tau$ is the characteristic decay time of the dynamics, $\beta$ is the so-called stretching exponent. The results of least-squares-fits to the data for $\tau$ are summarized in **Figure 1d**. The value of $\beta$ scatters around 1.1 (1.1 ± 0.1). For decreasing temperatures the decay time, $\tau$, grows down to 100 K in the yellow region, by more than a factor ten. Then $\tau$ stays fairly constant in the 50-100 K range (grey region) and decreases again below 50K. Thus, we can clearly distinguish *quasi-static* (or 'strongly pinned') CDW puddles, namely hard CDW in the 50-100 K range from more *dynamical* (or 'weakly pinned') CDW puddles, namely soft CDW, at both higher (T>100 K) and lower (T<50 K) temperatures.

In order to shed light on the correlation between the decay time, $\tau$, and the in-plane correlation length, $\xi_a$ we have plotted $\tau$, and, $\xi_a$, as a function of the temperature in **Figure 2a**. The in-plane coherence length is given by $\xi_a = 1/2\pi\Delta(H)$ where $\Delta(H)$ is the full width at half maximum of the CDW peak (see Methods). The data show that *soft* CDW (small $\tau$) occur in the presence of shorter correlation lengths (small $\xi_a$) at both low and high temperatures. It is known that the relation between the fluctuation times, temperature and correlation length can be described by the *activated dynamical scaling* (ADS) model [21,22]. In this model the competition between different interactions is controlled by the distribution of the energy barriers, which typically evolve as a power of the length scale which characterize the process [22]. The model is appropriate to describe the glassy freezing of charge and spin stripes below 50 K as shown by measurements of temperature dependent dielectric responses [31-34]. In the ADS model the fluctuation time obeys to $\tau \sim \exp(C\xi^z/T)$ where $\xi$ is the correlation length, and z the so-called dynamical critical exponent, which is typically around 2 and C is a constant. The term $C\xi^z$ times $k_B$, the Boltzmann constant, is an effective energy barrier height for activated fluctuations. We have developed our computer code for the ADS model introducing a constant correlation length, $\xi_a^0$, to take structural defects of the sample into account. The dashed line in Figure 3a shows the fitting curve to the measured characteristic decay times τ obtained by the modified ADS model

$$\tau = \tau_0 e^{C(\xi_a - \xi_a^0)^z/T} + \tau_1 \tag{3}$$

with $\xi_a^0$ = 7.56 nm, z = 2.2, and $\tau_1$ is an offset. The ADS model allows us to fit our experimental data fairly well in **Figure 2b** where we plot the decay times, τ, and coherence lengths, $\xi_a$, as a function of the incommensurability, η=14/H, where H is the CDW wavevector along the a* direction (see Methods). One key result of this work is that the *soft* CDW signal with short coherence lengths occurs both with lower and higher incommensurability values. The CDW behaviour alongside the ADS model has been compared with SDW behaviour, described in [20], in **Figure 2c**. We note that the SDW puddles are larger and exhibit a larger coherence length, which result typically in slower fluctuations with larger characteristic decay times.

In order to unveil the differences between the spatial inhomogeneity of CDW in nickelates and the inhomogeneity observed in cuprates [10], we used SµXRD to probe the local charge order via the mapping of the CDW superlattice peak. Above, we have shown that there are two types of CDW in the nickelates: fast fluctuating CDW puddles with short coherence length, *soft CDW*, and slower fluctuating CDW puddles with longer coherence length, *hard CDW*. Therefore, we investigated how these two types of CDW are distributed in space, focusing on the low temperature region where quantum fluctuations are expected for domain walls in incommensurate CDW.

We performed SµXRD measurements on the P10 beamline of the PETRA III synchrotron. The spatial distribution of the CDW peak incommensurability, η, [19,35] and the in-plane coherence length, $\xi_a$, have been calculated as described in Methods, and visualized over areas of 66×100 µm$^2$ in steps of 2 µm in both directions. Maps collected by scanning over the same sample area in the low temperature region at T=30 K, T=50 K and T=65 K are shown in **Figure 3a**. In the spatial maps, red (blue) areas correspond to a higher (lower) CDW peak incommensurability and coherence length. We clearly observe a different texture made of larger domains in the coherence length maps. **Figure 3b** shows a statistical analysis of the CDW incommensurability and coherence length spatial distribution in terms of the probability density function, PDF. At all temperatures, the incommensurability distributions follow a narrow normal distribution, attesting the good quality of our sample. The PDF are normally distributed but fluctuations of coherence length are larger over a factor 100, respect to the incommensurability fluctuations. This is confirmed by the spatial intensity correlation function, G(r) (**Figure 3c**), where r=|$R_i$-$R_j$| is the distance between x–y positions $R_k$ on the sample, G(r) function shows that the spatial correlation of incommensurability decay quite fast inside the resolution limited distance of 2 µm, while the spatial correlations of coherence lengths decay slower, outside distances of r = 10 µm. Details on the G(r) calculations have been described elsewhere in [44].

In **Figure 3d** we show the reconstructed map for the decay rate, given by $\tau^{-1}$, calculated point by point using the ADS model of Eq. 3, at each temperature T=35 K, 50 K and 65 K. The spatial inhomogeneity of the CDW rate fluctuations, $\tau^{-1}$, has been quantified by studying the connectivity in the maps of $\tau^{-1}$ using standard cluster 2D percolation analysis [50]. We consider two adjoining pixels to belong to the same cluster if they are connected along the horizontal, vertical, or diagonal direction and have rate below a threshold values $(\tau^{-1})^*$. We have calculated all the forming clusters, picking out the cluster with the largest extent, as a function of $(\tau^{-1})^*$. When we find a spanning cluster, with size equal to the system size, the system percolates. Thus in each map of $\tau^{-1}$, measured at the three different temperatures in the low temperature regime, we have calculated the percolation threshold, $p(\tau^{-1})$ the spanning cluster size and the number of clusters formed. The results are shown in **Figure 3e** and **Figure 3f**. At the lowest temperature, T=35K, we have found the largest percolation threshold of the $\tau^{-1}$ rate as well as the smallest spanning cluster. In addition, the number of *soft clusters* is larger at 50K and 65K.

**Conclusions**

We have investigated the spatial distribution of the charge density wave fluctuations rate in La$_{1.72}$Sr$_{0.28}$NiO$_4$ in the low temperature regime for T<65K. We have combined scanning micro X-ray diffraction (SµXRD) with resonant soft X-ray photon correlation spectroscopy (XPCS) to get spatial and temporal correlation landscapes of CDW textures. We report here clear evidence for CDW motion in nickelates in contrast with CDW remarkable stability in cuprates. The CDW fluctuations rate in nickelates increases with decreasing CDW coherence length in agreement with spin order dynamics [20]. We have found the anomalous drop at low temperatures of the CDW coherence length with the increasing fluctuation rate. We identify soft CDW puddles characterized by smaller size and higher mobility and hard CDW puddles with larger size and low mobility. Cluster analysis of SµXRD spatial maps of coherence lengths shows a phase separation between percolating soft CDW dynamic puddles and hard CDW puddles. The visualization of both spatial and time inhomogeneous landscape of charge density wave in $La_{2-x}Sr_xNiO_4$ provides a novel nanoscale phase separation phenomenon in complex quantum materials, supporting the proposals that this nanoscale phase separation involves lattice degrees of freedom.

**Methods**.

Single-crystalline $La_{1.72}Sr_{0.28}NiO_4$ was grown by floating zone technique. The seed and feed rods were prepared from polycrystalline powder obtained by solid state reaction of $La_2O_3$, $SrCO_3$ with an excess of $NiO$. The reaction was performed at 1200 °C for 20 h with intermediate grinding. The rods were densified at 1500 °C for 5 h. The synthesis were carried out in air.

The temperature dependent scanning micro X-ray diffraction µXRD experiments were carried out at the Coherence Beamline P10 of PETRA III synchrotron Hamburg. The synchrotron radiation source was a 5 m long undulator (U29). The x-ray beam was monochromatized by a cooled Si(111) double-crystal monochromator with a bandwidth of $\Delta E/E \sim 1.4 \times 10^{-4}$. The collimated coherent x-ray beam was focused using a beryllium refractive lens (CRL) transfocator to a size of about $2 \times 2.5$ µm$^2$ at the sample positioned ~1.6 m downstream of the transfocator center. The incident flux on the sample was about $10^{11}$ photons/s. The windows of the He cryostat as well as the entrance window of the evacuated detector flight path were covered by 25 µm thick Kapton foils. The incident photon energy was set slightly below the Cu $K$-edge to minimize a possible fluorescence background. The scattered signal was detected using the large horizontal scattering setup with a sample-to-detector distance of 5 m. A PILATUS 300 K detector was used to record the x rays scattered by the sample. For the measurements, the sample was cooled to the lowest temperature 30K and the measurements were performed during a heating cycle. We aligned the crystal to detect the charge ordering satellite of the 100 reflection, which appears below $T_\xi$=120 K. The incommensurate charge density wave satellite appears at $q_h=2\varepsilon=0.59$. The scanning maps shown in Fig.3 have been obtained by translating the sample in steps of 2 µm in both directions. We have mapped the spatial distribution of the $(q_h,0,1)$ CDW peak over areas of $100 \times 66$ µm$^2$.

The CDW incommensurability is measured by the lattice units of the superstructure η related to its period, 1/H. The incommensurate CDW with 0.25<ε<0.33 consists of the mixture of the ε=1/n and the ε=1/m order, i.e alternating the so-called n=3 stripes portions and m=4 stripes portions. Indeed, quasi-commensurate periods $\lambda = 1/\varepsilon = \frac{n\,x+m\,y}{n+m}$ intermediate between two main commensurate wave-vectors 1/n and 1/m with periods of m and n lattice units, respectively, occur at integer numbers of lattice unit cells $\eta = \frac{n+m}{\varepsilon}$. Each quasi-commensurate phase (QCP) corresponds to a modulation wave locked with the underline lattice onto a rational number. This sequence of quasi-commensurate phases is called the Devil's staircase [17-19]. Choosing n=2x3 and m=2x4 the incommensurability in our sample is given by η=14/H which approaches the quasi-commensurate phase with η =24 lattice units around T=65K.

The CDW coherence length has been extracted from the full widths at half maxima, $\Delta(H)$, $\Delta(L)$ by fitting the profiles along H and L directions with Lorentzian-squared line shapes. In particular, the coherence length along the **a** (in-plane) and **c** (out of plane) crystallographic directions have been calculated as $\xi_a = 1/2\pi\Delta(H)$ and $\xi_c = 1/2\pi\Delta(L)$.

X-ray photon correlation spectroscopy (XPCS) measurements were carried out at beamline 12.0.2 of the Advanced Light Source at Lawrence Berkeley National Laboratory (USA). The experiment was conducted in a θ-2θ reflection geometry. Tuning the energy of the incoming linear σ polarized x-ray to the $L_3$-edge of Ni (852 eV) brings about magnetic sensitivity. The transverse coherence of the beam was established by placing a 5 microns pinhole approximately 3 mm in front of the sample. A CCD placed 0.45 m away served as detector.

**Acknowledgments**

A.R. and G.C. acknowledge the Stephenson Distinguished Visitor Program by DESY We are grateful to Nicola Poccia for experimental help. A.A.N. acknowledges funding from Ministry of Research, Technology and Higher Education through Hibah WCU-ITB. Support by Superstripes-onlus and by the DFG within SFB 925 – project 170620586 is gratefully acknowledged. The Advanced Light Source is a U.S DOE Office of Science User Facility under contract no. DE-AC02-05CH11231. We acknowledge DESY (Hamburg, Germany), a member of the Helmholtz Association HGF, for the provision of experimental facilities. Parts of this research were carried out at the P10 beamline of PETRA III.

**Author Contributions**

Each author made substantial contributions to the conception or design of the work; or the acquisition, analysis, or interpretation of data; or the creation of new software used in the work; or have drafted the work or substantively revised it. A.R. and G.C. conceived the project and designed the experiments; M.S. contributed to the planning of the experiments; the samples were grown by A.A.N.; micro X-ray diffraction experiments have been carried out at P10-PETRA III (DESY) by A.R., G.C., B.J., A.Z. and M.S.; the XPCS experiments at the Avanced Light Source have been carried out by S.M., S.R., and A.R.; data analysis has been done by G.C., A.B. and A.R.; the manuscript has been written by G.C., A.B. and A.R., collecting feedback from all the authors. Each author has agreed both to be personally accountable for the author's own contributions and to ensure that questions related to the accuracy or integrity of any part of the work, even ones in which the author was not personally involved, are appropriately investigated. Each author has approved the submitted version.

**Data availability statement:** all data generated or analyzed during this study are included in this published article. Further information on the current study are available from the corresponding author on reasonable request.

**Corresponding author:** Gaetano Campi, e-mail: gaetano.campi@ic.cnr.it

**Declaration of interests.** The authors declare no competing interest.

**Figures and figure captions**

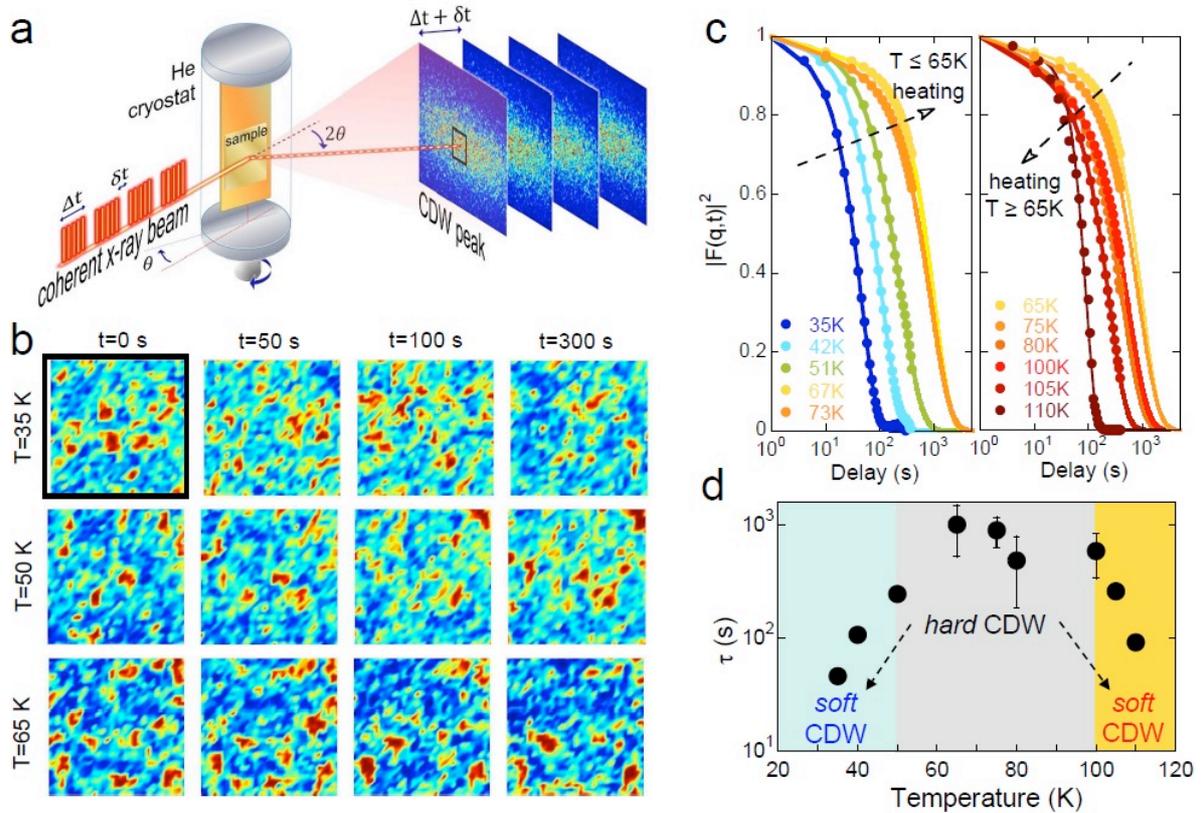

**Figure 1. Temperature dependence of CDW dynamics measured by resonant soft X-ray Photon Correlation Spectroscopy in La$_{1.71}$Sr$_{0.29}$NiO$_4$**

**(a)** Experimental setup for the XPCS measurements. Coherent X rays hit the sample placed at the 2θ angle of the CDW peak. A time series of X-ray coherent diffraction images allows to study the development of the speckle spatial correlations. The region of interest, ROI, inside the square, where the intensity is more intense and stable has been used for analysis. **(b)** Images of speckle data recorded at T=35K, T=50K and T=65K in the selected ROI at different times. The larger the delay time, the more the speckle pattern differs from the initial one. **(c)** Autocorrelation function vs. delay measured at the indicated different temperatures. Experimental data are indicated by full circles along the solid lines fits obtained by using the stretched exponential model of Eq. 2. **(d)** Time delay, τ, extracted by fitting the |F(q,t)|$^2$ using Eq. 2. The time delay of fluctuations reaches values higher by a factor 100 passing from the *hard* region to the *soft* regions at both high and low temperature.

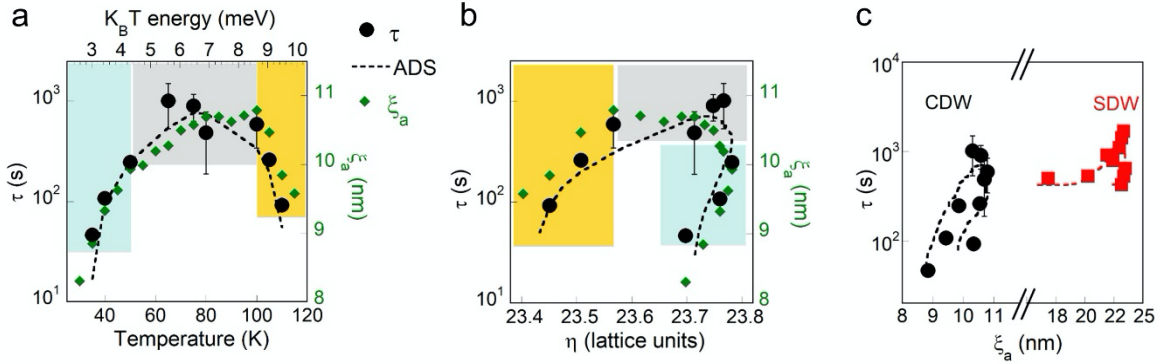

**Figure 2. Correlations between CDW and SDW domain size and dynamics in La$_{1.71}$Sr$_{0.29}$NiO$_4$**
**(a)** Characteristic decay times, $\tau$, of CDW (black full circles) along the fit (dashed line) extracted in the modified Activated Dynamical Scaling (ADS) model described by Eq. 3, as a function of temperature in logarithmic scale. The behavior of $\tau$ is compared with the (full diamonds) in plane coherence length, $\xi_a$. We observe how lower delays correspond to lower coherence lengths that is to smaller domains in both low temperature (light blue) and high temperature (yellow) regions. **(b)** Time delay, $\tau$, (black full circles), and $\xi_a$ (full diamonds) of CDW, as a function of the incommensurability, $\eta=14/H$, where H is the CDW wavevector along the a* direction (see Methods). Here lower values of delays and coherence lengths occur in the low temperature (light blue) and high temperature (yellow) regions. The fit of $\tau$ as a function of T and $\xi_a$ extracted from the ADS theory is represented by the black dashed line. **(c)** The correspondence between time delays, $\tau$, and domain size, $\xi_a$, for CDW has been compared with time delays and domain size in SDW measured in the same sample, as reported in ref. 20.

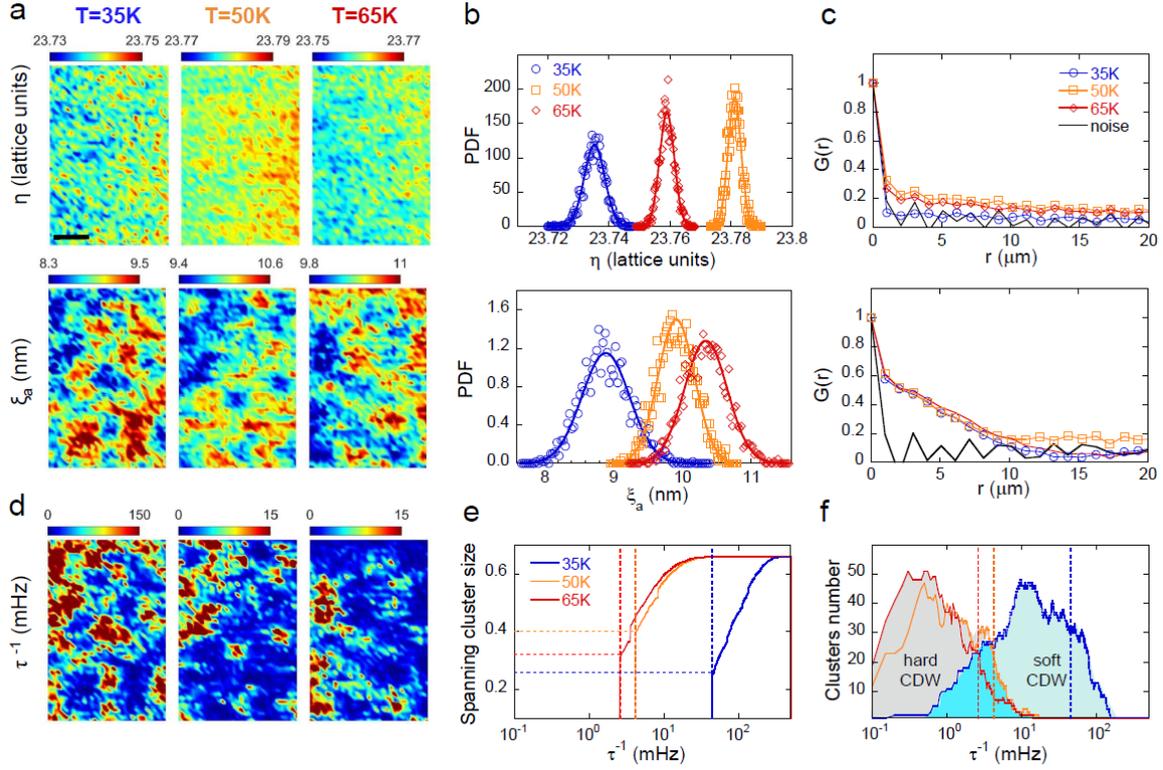

**Figure 3. Mapping the small/large and fast dynamics CDW domains by X-ray micro diffraction in La$_{1.71}$Sr$_{0.29}$NiO$_4$.**
(**a**) Maps (**b**) PDF and (**c**) G(r) of CDW peak incommensurability, η, in plane coherence length, ξ$_a$, measured at T=35K, 50K, and 65K. Values of each single pixel have been obtained recording the CDW peak in a specific (x, y) position of the sample. In order to reconstruct the spatial maps, the sample has been scanned over an area of about 66×100 μm$^2$ in steps of 2μm in both directions. The scale bar shown in the upper frame corresponds to 20μm. We can observe the narrow fluctuations of η that decays in few microns. On the other hand, ξ$_a$ form puddles of about 10μm of radius. (**d**) Maps of CDW peak frequency, τ$^{-1}$, at T=35K, 50K and 65K. (**e**) Spanning clusters size and (**f**) forming clusters number calculated for the maps in (a). The percolation thresholds are represented by the vertical dashed lines.